\documentclass[trackchanges,twocolumn]{aastex7}
\usepackage{amsmath}
\usepackage{subcaption} 

\newcommand{\comment}[1]{}
\usepackage{bm}

\newcommand{\Addp}[1]{\textcolor{black}{#1}}

\newcommand{\Erase}[1]{}

\newcommand{\M}{{\ $\mid$}}

\newcommand{\W}{{$\lambda$}}

\newcommand{\q}{$q_{\rm ion}$}

\newcommand{\K}{{\rm K}}

\begin{document}
\title{
A Black-Hole Envelope Interpretation for Cosmological Demographics of Little Red Dots
}

\correspondingauthor{Hiroya Umeda}

\author[0009-0008-0167-5129]{Hiroya Umeda}
\affiliation{Institute for Cosmic Ray Research,
The University of Tokyo,
5-1-5 Kashiwanoha, Kashiwa,
Chiba 277-8582, Japan}
\affiliation{Department of Physics, Graduate School of Science, The University of Tokyo, 7-3-1 Hongo, Bunkyo, Tokyo 113-0033, Japan}
\email[show]{ume@icrr.u-tokyo.ac.jp}

\author[orcid=0000-0000-0000-0001]{Kohei Inayoshi}
\affiliation{Kavli Institute for Astronomy and Astrophysics, Peking University, Beijing 100871, China}
\email{inayoshi@pku.edu.cn}  

\author[orcid=0000-0002-6047-430X]{Yuichi Harikane}
\affiliation{Institute for Cosmic Ray Research, The University of Tokyo, 5-1-5 Kashiwanoha, Kashiwa, Chiba 277-8582, Japan}
\email{hari@icrr.u-tokyo.ac.jp}

\author[orcid=0000-0002-5358-5642]{Kohta Murase}
\affiliation{Department of Physics; Department of Astronomy \& Astrophysics; Center for Multimessenger Astrophysics, Institute for Gravitation and the Cosmos, The Pennsylvania State University, University Park, PA 16802, USA}
\affiliation{Center for Gravitational Physics and Quantum Information, Yukawa Institute for Theoretical Physics, Kyoto University, Kyoto, Kyoto, 606-8502, Japan}
\email{murase@psu.edu}

\begin{abstract}
%
Little red dots (LRDs) newly discovered with JWST are active galactic nuclei (AGN) that may represent black hole (BH) growth at the earliest cosmic epochs.
These sources show puzzling features unlike typical AGNs, 
including red optical continua, weak hot-dust emission, and a lack of detectable X-rays. 
Previously, LRDs have often been interpreted as dust-reddened AGNs, 
leading to severe inconsistencies with the luminosity and BH mass densities 
inferred for previously known AGNs over $0<z<5$. 
The BH-envelope (BHE) model has been proposed to explain these characteristics, in which an accreting BH is enshrouded by a dense, optically thick gaseous envelope.
In this Letter, we reanalyze the SEDs of $\sim 400$ photometric LRDs in the COSMOS-Web survey using the BHE model and reassess their implications for cosmological BH evolution. 
We find that the optical-NIR spectra of LRDs are well reproduced by blackbody emission 
with an effective temperatures of $4000-6000~\K$. 
Within the BHE framework, the inferred bolometric luminosities decrease by 
$\gtrsim1-2$ orders of magnitude compared to dust-reddened AGN assumptions.
As a result, the revised luminosity function, BH accretion density, and 
BH mass function become consistent with those of AGNs at $z<5$.
The stellar masses of LRD hosts are estimated by attributing the UV excesses to star formation.
Although the resulting $M_{\rm BH}/M_\star$ ratio remains higher than the local empirical value, the excess is modest. 
Overall, the BHE model not only resolves the spectral features of LRDs but also brings their statistical properties into agreement with the broader cosmological BH population.
\end{abstract}



\section{Introduction} \label{intro}
JWST has uncovered a population of compact, red nuclei with broad emission features, known as little red dots \citep[LRD; e.g., ][]{Matthee24,Harikane23b,Greene24,Kocevski23,Akins25}. Their rest-frame UV-optical spectra show a distinctive V-shaped continuum with an inflection point near the Balmer limit wavelength and a smooth rise toward longer wavelengths \citep{Greene24,B.Wang24b,Setton24,Kocevski25,Labbe25}. Their spectral energetics, compact morphologies, and broad lines indicate active black hole accretion at high redshifts \citep{Matthee24,Hviding25}, though their intrinsic luminosities and masses remain uncertain due to the poorly constrained spectral origin. While dust-reddened quasars can reproduce their red optical colors, they struggle to account for the weakness of near-to-mid infrared emission constrained by JWST/MIRI \citep{Williams24,Perez-Gonzalez24,Akins25}. Moreover, the absence of X-ray emission further implies Compton thick, super-Eddington accretion flows \citep{Maiolino25_Xray,Madau_Haardt24,Inayoshi25_Xray}, for which the bolometric corrections are largely unconstrained even for nearby AGNs.

A growing number of observations and theoretical studies suggest that LRDs are powered by accreting black holes enshrouded by a dense, optically thick envelope \citep[e.g.,][]{Inayoshi25a,Nadiu25,DeGraaff25,Lin25,Inayoshi25b,Kido25}. This envelope forms a photosphere that thermalizes a large fraction of the accretion power and produces a blackbody-like continuum with an effective temperature of $\sim 5000~\K$ \citep[e.g.,][]{Hayashi1961}. The presence of prominent Balmer absorption and break features reinforce this interpretation because hydrogen-rich photospheres at such temperatures naturally imprint these features \citep{Liu25}.
Theoretical models for super-Eddington flows and direct-collapse seed formation predict radiation-pressure supported envelopes and outflows that can trap high-energy photons and reprocess the emission to optical wavelengths (\citealt{Kido25,Begelman25}; see also \citealt{Begelman06,Hosokawa13}).

Motivated by these recent developments in a context of LRDs, we revisit the LRD population with a black-hole-envelope model. We fit the optical bumps as photospheric emission and we quantify the statistical properties of the envelopes and the central black holes. Our analysis aims to place LRDs within a physically motivated growth pathway for early accreting black holes and to clarify how their demographics connect to super-Eddington accretion at high redshift.

Throughout this Letter, we assume a flat $\Lambda$ cold dark matter (CDM) cosmology consistent with the constraints from Planck \citep{Planck}: $h=0.6766$, $\Omega_{\rm m}=0.3103$, $\Omega_\Lambda=0.6897$, and  $\Omega_{\rm b} h^2=0.02234$. 
All magnitudes are in the AB system \citep{1983ApJ...266..713O}.
\section{Data and SED Fitting}
\subsection{Little Red Dot Sample}

In this work, we utilize the LRD sample selected from the COSMOS-Web survey with a total 0.54 deg${}^2$ area \citep[GO-1727; PI: Jeyhan Kartaltepe;][]{Casey23}
imaging data by \cite{Akins25}\footnote{\url{https://hollisakins.com/projects/1_project/}}.
The selection requires a red optical color with $m_{277}-m_{444}>1.5$, 
and a compact morphology in the F444W-band image. \cite{Barro24} found that almost all the objects selected by this criterion are compact and appear blue at 1 to 2 $\mu m$.
The compactness is quantified by the flux ratio measured with two aperture sizes, i.e., $0.5<C_{444}<0.7$, where $C_{444}=f_{444}(d = 0.2\arcsec)/ f_{444}(d = 0.5\arcsec)$. 
This criterion selects the objects with small effective radii (i.e., $r_e\lesssim 100-300$ pc). The full LRD sample size consists of 434 objects.

To improve the inference of SED at the rest-frame optical wavelengths, we additionally select objects with MIRI band photometry to ensure multiple photometry data to constrain the optical spectrum shape. This fiducial LRD sample with MIRI data coverage consists of 148 objects. While we conduct the SED fitting for full LRD sample, we only use the fiducial sample to measure the bolometric luminosity function.

\subsection{Black Hole Envelope Model Fitting}
\citet{Akins25} performed SED fitting on the LRD sample to estimate their photometric redshifts and bolometric luminosities. They adopted quasar template spectra with dust attenuation, and found that the LRDs are located at photometric redshifts of $z\sim5-9$. However, the non-detection of the rest-frame NIR in their multi-wavelength stacking analysis of the LRDs suggests a lack of dust re-emission after attenuation. Moreover, the stacking analysis of LRDs suggests non-detection of X-rays, contradicting with the observational characteristics of ordinary AGN with super-Eddington accretion. To reconcile the puzzling features of LRDs, \cite{Kido25} propose the Black Hole Envelope (BHE) model. In the BHE model, intense outflows from the central black hole are confined within a massive envelope. The energy from the central engine is trapped and convected within the envelope, eventually escaping from its photosphere by blackbody radiation, similar to stars evolving along the Hayashi track. Because the photospheric radiation in the BHE model is expected to be thermal emission with an effective temperature of $\sim4000-7000$ K, its Wien tail naturally reproduces the optical V-shaped continuum while simultaneously explaining the non-detection of rest-frame NIR emission by JWST/MIRI \citep{Inayoshi25c}. Motivated by the convenience of the model, we fit the SED of LRD sample with the BHE model to estimate the photometric redshift and bolometric luminosity of the object. \par

\begin{deluxetable*}{cccccccccc}
  \tablecolumns{10}
  \tabletypesize{\scriptsize}
  \tablecaption{Prior Distributions for Fitting Parameters%
  \label{table:prior}}
  \tablehead{%
  \colhead{$z$} &
  \colhead{$T_{\rm ph}$} &
  \colhead{$\beta_{\rm UV}$} &
  \colhead{$A_1$} &
  \colhead{$A_2$} &
  \colhead{$A_V$} \\
  \colhead{} &
  \colhead{($\times10^3$~K)} &
  \colhead{} &
  \colhead{(nJy)} &
  \colhead{(nJy)} &
  \colhead{(mag)} \\
  \colhead{(1)} &
  \colhead{(2)} &
  \colhead{(3)} &
  \colhead{(4)} &
  \colhead{(5)} &
  \colhead{(6)} &
  }
  \startdata
  U(0,10) & 
  $\rm N_{t}$(5, 1, -4, 4) & 
  N(-2.5,0.5) & 
  U(0,$[f+\sigma]_{\rm F444W}$) & 
  $\rm\ln N$($\ln A_1$,$\ln 2$) & 
  $\rm U$(0,3)
  \enddata
  \tablecomments{
  (1): Systemic redshift. 
  (2): Temperature of blackbody radiation from the black hole envelope. 
  (3): Slope index for power-law UV component. 
  (4): Amplitude of blackbody radiation at Balmer break. Here, $[f+\sigma]_{\rm F444W}$ represent the $1\sigma$ upper error values for the observed F444W flux.
  (5): Amplitude of power-law UV component at Balmer break.
  (6): Visual extinction.\\
  N($\mu$, $\sigma)$ is a normal distribution with mean $\mu$ and variance $\sigma^2$. $N_{\rm t}$($\mu$, $\sigma, n_l, n_u)$ is a truncated normal distribution with mean $\mu$ and variance $\sigma^2$, with truncation below (above) $n_l(n_u)\times \sigma$. U($a$, $b$) is a uniform distribution between $a$ and $b$. $\rm \ln N$($\mu$, $\sigma)$ is a log-normal distribution with mean $\mu$ and variance $\sigma^2$.}
  \end{deluxetable*}

For the BHE model, we approximate the optical spectrum of LRDs by blackbody radiation with a corresponding photosphere temperature $T_{\rm ph}$. We add the break feature to the blackbody radiation at the Balmer break wavelength $\lambda_{\rm Balmer}$ (i.e., 3646~{\AA}) to account for the Balmer absorption within the BHE. We also allow moderate dust attenuation on the blackbody radiation by changing visual extinction value ($A_V$). For the dust attenuation law, we assume \cite{2000ApJ...533..682C}. The optical part of the BHE model spectrum is expressed as follows:
\begin{equation}
f_\nu \propto H(x) B_\nu(T_{\rm ph}) e^{-\tau_\nu(A_V)},
\label{bhe_opt_eq}
\end{equation}
where $x=\nu-c/\lambda_{\rm Balmer}$ and $H(x)$ is the Heaviside step function. We normalize the optical part of the spectrum (i.e., equation \ref{bhe_opt_eq}) to $A_1$ at the Balmer break in the observed-frame.
For the full spectrum model, we add a power-law spectrum with the slope $\beta_{\rm UV}$ to equation \ref{bhe_opt_eq} account for the UV component of LRD spectrum. We also normalize the power-law component to $A_2$ at the Balmer break in the observed-frame. In addition, we add a break at the rest-frame 1216~{\AA} to account for the Lyman break. 
We fit the spectra of the LRD sample with the SED model described above. We use the Markov Chain Monte Carlo (MCMC) module \texttt{PyMC} \citep{2024zndo...4603970P} to perform the SED fitting for the flexible sampling. We set conventional prior distributions for each parameters as described in Table \ref{table:prior}. We briefly explain key details on the physically and observationally motivated reasoning behind our prior selections. We vary $T_{\rm ph}$ around $3000-7000~$K to match the typical photosphere temperature of BHE model described in \cite{Kido25}. We also model $A_1$ and $A_2$ with mutual dependencies, where $A_2$ is tied to $A_1$ to ensure a continuous transition across $\lambda_{\rm Balmer}$ and to reproduce the observed V-shaped spectral feature \citep[e.g.,][]{Nadiu25,DeGraaff25}. We also limit $A_1$ value smaller than the $1\sigma$ upper error of the observed f444W flux because the selection of \cite{Akins25} requires the elevated rest-frame optical (i.e., blackbody) component compared to the UV component. We also vary $A_{\rm V}$ around 0 to 3 to account for relatively weak dust continuum emission in the infrared \citep[e.g.,][]{Akins25}. For the photometric data, we use the all available photometry provided in the \cite{Akins25}. For the non-detection band, we utilize the 2$\sigma$ upper limit and create the likelihood function following the method in \cite{Sawicki12}. To estimate the detection limit in each band, we fitted a linear relation between the signal-to-noise ratio and the flux uncertainty provided in the original catalog of \citet{Akins25}. We use all available photometry provided in the catalog from the HST/Advanced Camera for Surveys (ACS), JWST/
NIRCam, and JWST/MIRI imaging. From the COSMOS-Web data, we use (HST)/ACS F814W and JWST/NIRCam F115W, F150W, F277W, F444W, and JWST/MIRI F770W filters. For sources covered by the PRIMER survey \cite[GO-1837; PI. James Dunlop;][]{2021jwst.prop.1837D}, we additionally include photometry from the following filters: HST/ACS F666W and F814W, JWST/NIRCam F090W, F115W, F150W, F200W, F277W, F356W, F410M, F444W, and JWST/MIRI F770W and F1800W. To reduce the impact of outliers caused by strong emission lines, we adopt a likelihood function that is robust to such deviations. Instead of the commonly used Gaussian likelihood, we employ a Student’s t likelihood, defined as:
\begin{equation}
  \begin{aligned}
    \phi(f_{i}^{\rm obs}|\vec{\theta})&=2\frac{\Gamma\left(\frac{\nu+1}{2}\right)}{\sqrt{\nu\pi}\sigma_i\Gamma\left(\frac{\nu}{2}\right)}\\
    &\quad \times \left(1+\frac{1}{\nu}\left(\frac{f_{i}^{\rm obs}-f_{i}^{\rm model}(\vec{\theta})}{\sigma_i}\right)^2\right)^{-\frac{\nu+1}{2}},
  \end{aligned}
\end{equation}
We adopt the degree of freedom $\nu=1$ for the Student's t distribution. We also adjust the non-detection term of the cumulative likelihood function based on the Student's t distribution.
\begin{equation}
  \Phi(f_{i,up}^{\rm obs}|\vec{\theta})=\int_{-\infty}^{f_{i}^{\rm obs,up}} \phi(x|\vec{\theta}) dx,
\end{equation}
where $f_{i,up}^{\rm obs}$ is the 3$\sigma$ upper limit flux for the non-detection band. The total likelihood function is expressed as follows:
\begin{equation}
  \mathcal{L}(\vec{\theta}|\vec{f^{\rm obs}})=\prod_{i}^{N_{\rm det}}\phi(f_{i}^{\rm obs}|\vec{\theta})\prod_{j}^{N_{\rm non-det}}\Phi(f_{j,up}^{\rm obs}|\vec{\theta}),
\end{equation}
We perform MCMC sampling with 20 parallel chains and 20,000 steps per chain.

\begin{figure*}[t!]
\centering
\begin{subfigure}{0.45\linewidth}
\includegraphics[width=\linewidth]{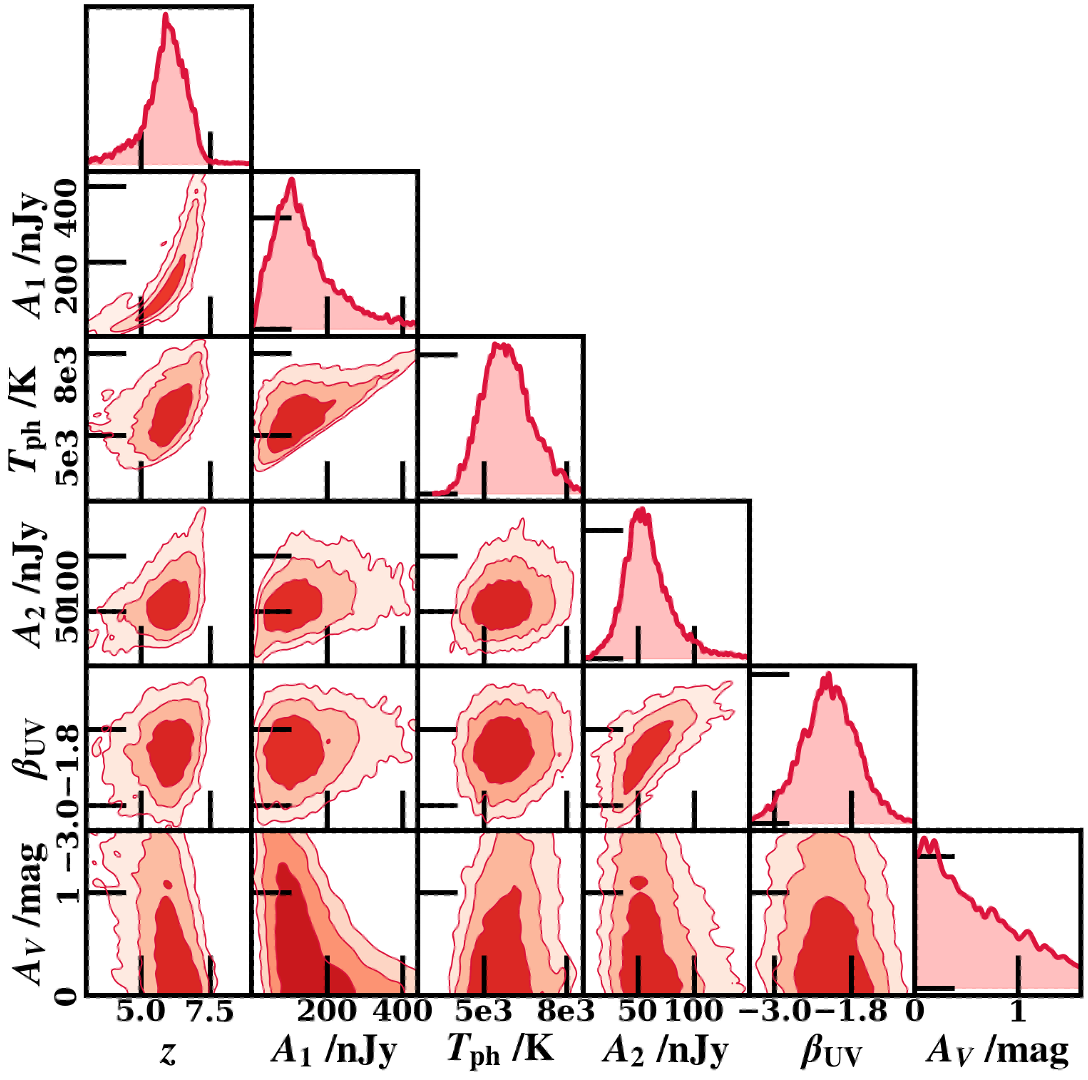}
\end{subfigure}\hfill
\begin{subfigure}{0.55\linewidth}
\includegraphics[width=\linewidth]{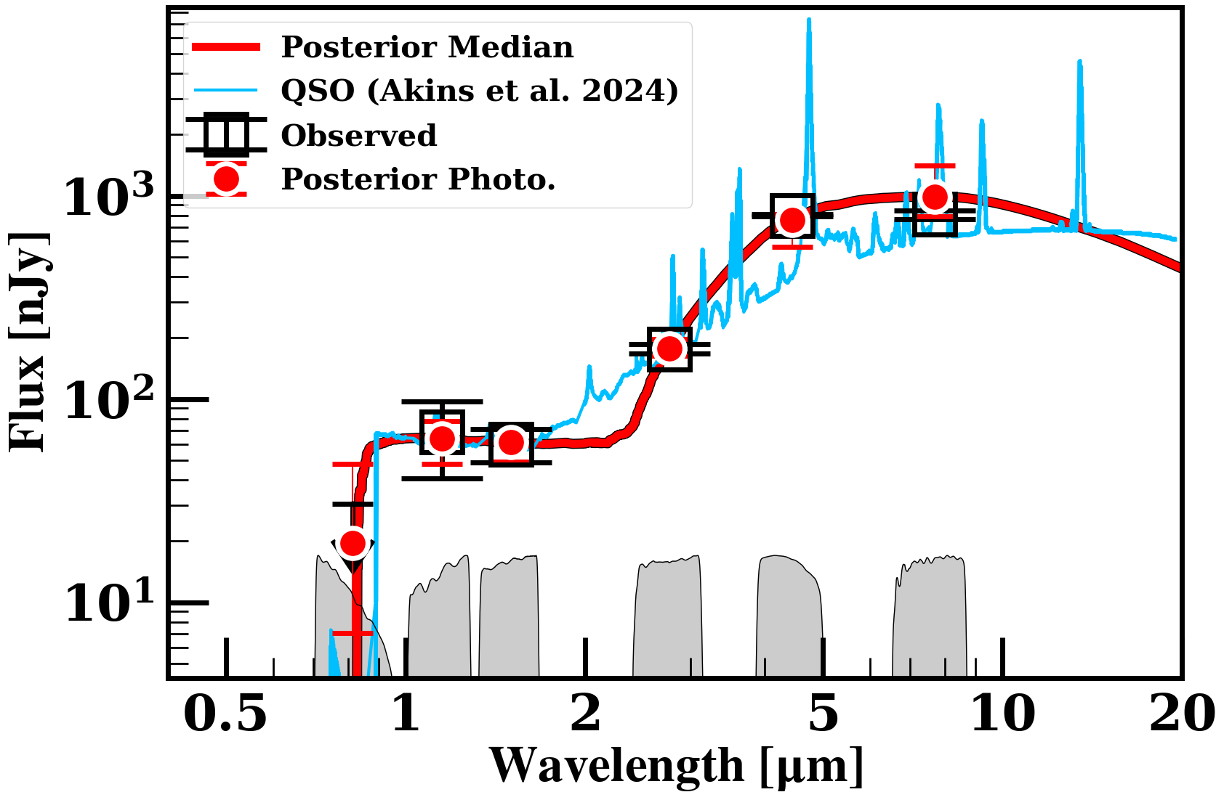}
\end{subfigure}
\caption{Left: Example posterior distribution from the SED fitting. 
The top panels in each column represent the 1D marginalized PDF for the corresponding parameters, while the remaining panels display the 2D marginalized PDFs for each parameter pair. Right: Posterior PDF of the BHE-based SED fitting for one LRD from \cite{Akins25} (COS-145575). Black data points with error bars show the observed photometric fluxes. The red solid lines indicate SED realization drawn from the posterior PDF, and the red points show the corresponding model photometric fluxes and uncertainties. The median and 68-th percentile parameter constraints are as follows: $z={6.00}^{+0.64}_{-0.77}$, $A_1/{\rm nJy}={124}_{-58}^{+101}$, $T_{\rm ph}/{\rm K}={5670}_{-690}^{+820}$, $A_2/{\rm nJy}={56.3}_{-16.5}^{+22.6}$, $\beta_{\rm UV}={-2.14}_{-0.44}^{+0.43}$, and $A_{V}/{\rm mag}={0.55}_{-0.40}^{+0.72}$. The shaded curves at the bottom of the panel represent the transmission curves of the filters used in the fitting.}
\vspace{3mm}
\label{fit_ex} 
\end{figure*}

\begin{deluxetable}{ccc}
  \tablecolumns{2}
  \tabletypesize{\scriptsize}
  \tablecaption{Summary of Posterior Inferences from BHE Modeling of the LRD Sample
  \label{table:prop}}
  \tablehead{
    \colhead{Quantity} & ~~~~~~~&\colhead{Value}
  }
  \startdata
  $T_{\rm ph}$ (K)        &                & $5420^{+240}_{-200}$ \\
  $A_V$ (mag)              &                    & $0.93^{+0.31}_{-0.27}$ \\
  $L_{\rm bol}/\lambda L_{\lambda,5100}$ & & $4.41^{+2.12}_{-1.37}$ \\
  \enddata
  \tablecomments{Values are medians with 16th and 84th percentile uncertainties for LRD sample.}
\end{deluxetable}

\subsection{Fitting Result}
In Figure \ref{fit_ex}, we present an example of the fitting for one LRD candidate (i.e., COS-145575 in \citealt{Akins25}). The left panel of Figure \ref{fit_ex} presents the corner diagram representing both the 1D and 2D marginalized posterior probability distributions. The median posterior spectrum obtained from our SED fitting is shown in the right panel. We also present the model photometry, calculated by convolving the model spectrum with the filter transmission curves. Our model and the observed photometry are consistent within $1\sigma$ error bars in all imaging bands. For comparison, we show the best-fit QSO spectrum from \citet{Akins25}. Unlike standard QSO templates, \cite{Akins25} introduce an additional unattenuated intrinsic component to reproduce the unusually blue UV continuum together with very red optical component. Their QSO model spectrum resembles an LRD spectrum, with enhanced optical flux due to strong emission lines and suppressed UV emission caused by dust attenuation. However, the QSO model would produce high rest-frame near-infrared flux values from the strong dust re-emission, which is hard to reconcile with the non-detections in near-infrared in the stacked spectra \citep{Akins25}.

\begin{figure*}[t]
\centering
\begin{subfigure}{0.5\linewidth}
\includegraphics[width=\linewidth]{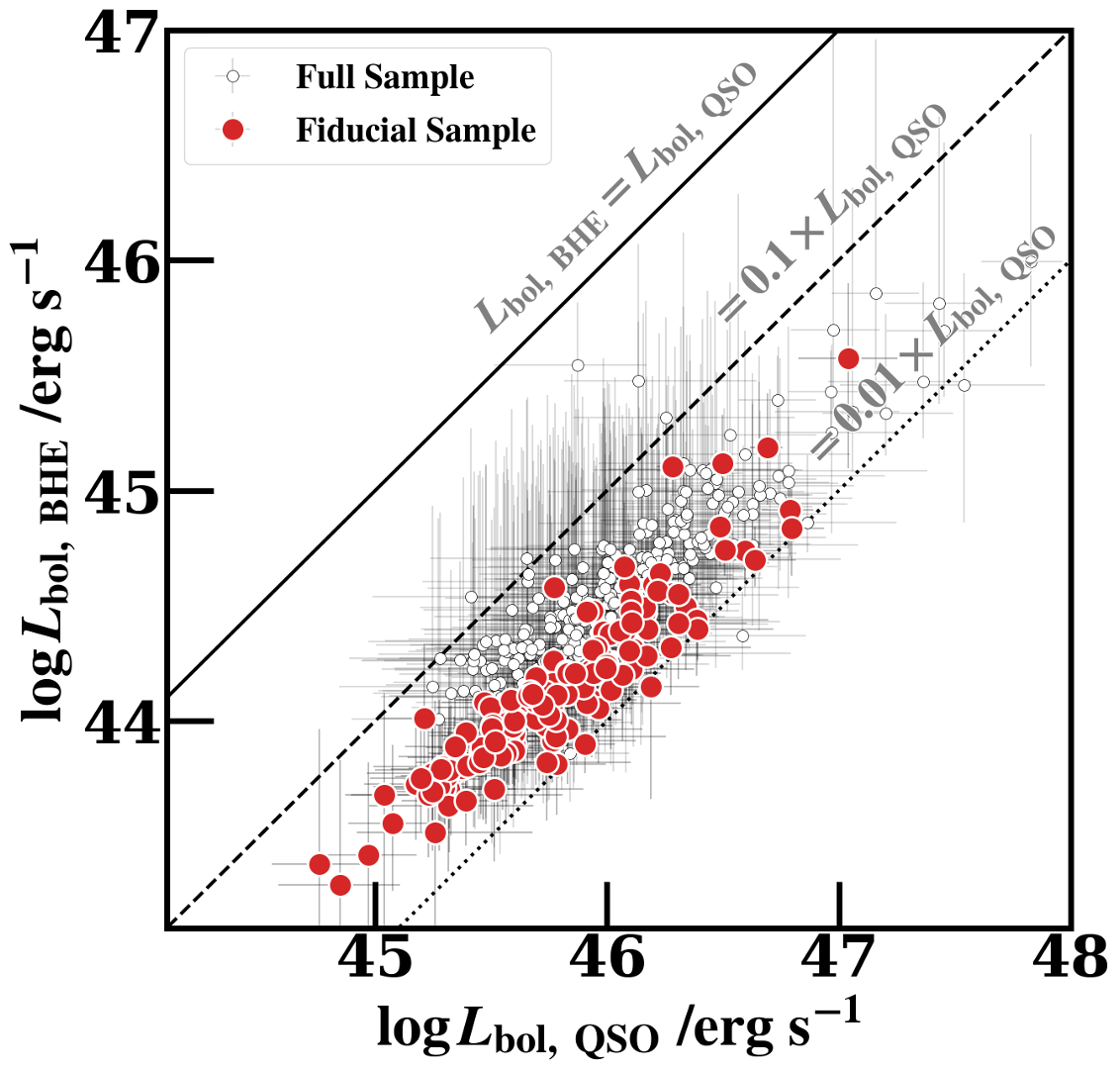}
\end{subfigure}\hfill
\begin{subfigure}{0.49\linewidth}
\includegraphics[width=\linewidth]{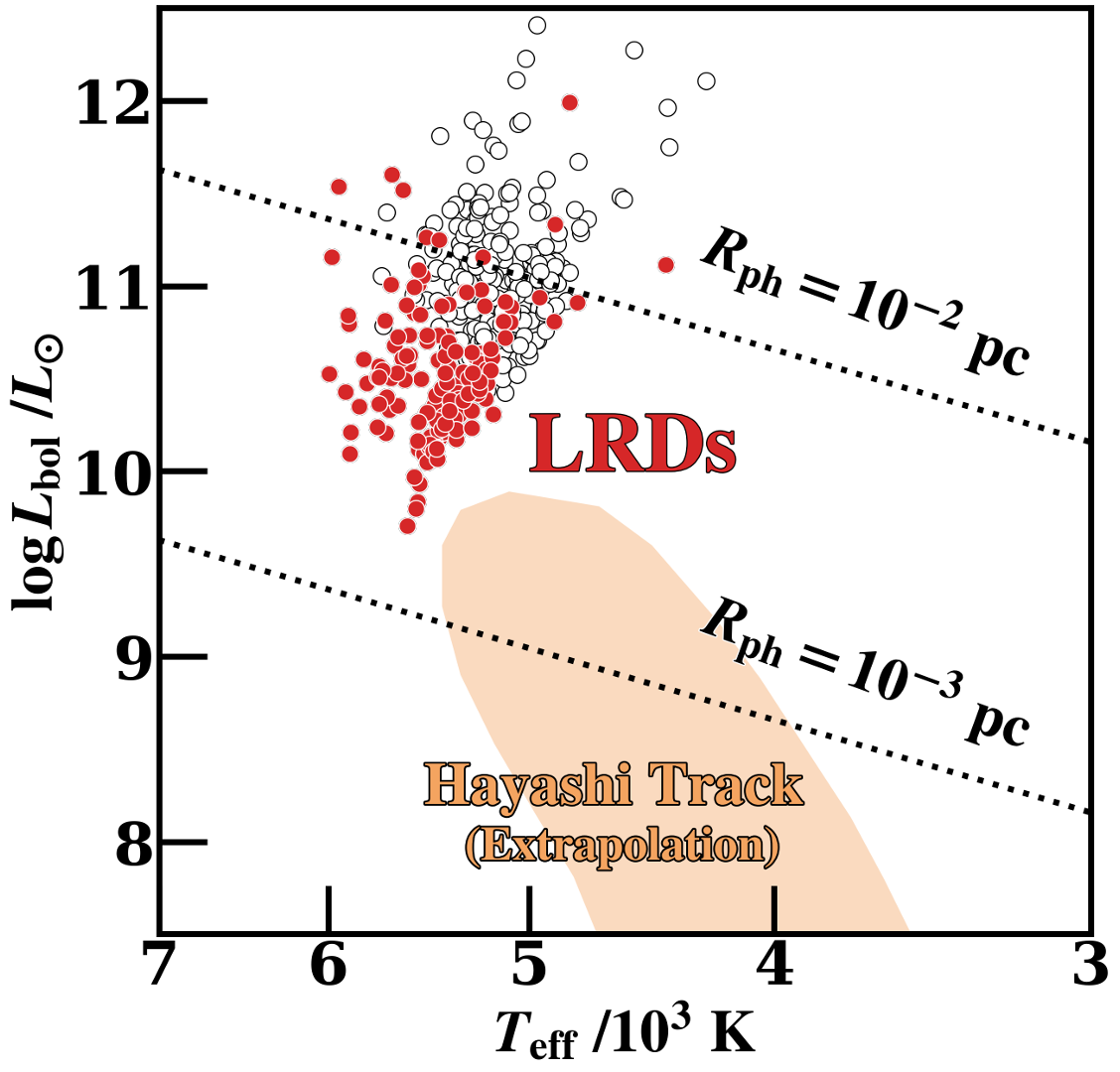}
\end{subfigure}
\caption{Left: Comparison of the bolometric luminosities between the BHE model ($L_{\rm bol,~BHE}$) and dust-reddened AGN model ($L_{\rm bol,~QSO}$).
The sold, dashed, and dotted lined represent $L_{\rm bol,~BHE}/L_{\rm bol,~QSO}=1.0$, $0.1$, and $0.01$, respectively. 
Right: Hertzsprung-Russell diagram for the fiducial LRD sample (red circles) and full sample (open circles), showing effective 
temperatures and luminosities under the assumption of blackbody spectra.
Dotted lines indicate the loci of constant photospheric radius of $R_{\rm ph}=10^{-3}$ and $10^{-2}~{\rm pc}$.
The narrow temperature range over $T_{\rm eff}\simeq 4500-6000~{\rm K}$ is consistent with the tip of the Hayashi track
extrapolated from from the locus of red supergiant stars in the Milk Way (shaded region).
}
\vspace{3mm}
\label{comparison} 
\end{figure*}

Table \ref{table:prop} summarizes the median posterior values for our sample. To further compare our inferences with previous work,
we directly compare the bolometric luminosity estimate for individual LRDs. We adopt the bolometric luminosities from \citet{Akins25} (i.e., $\log L_{\rm bol,~QSO}$), which they derived with their dusty QSO spectral model. For our BHE model, we compute the bolometric luminosity, $\log L_{\rm bol,~BHE}$, by integrating the intrinsic (unattenuated) blackbody spectrum. We do not include the UV power-law component into our bolometric luminosity measurement. 
As shown in the Figure \ref{comparison}, the bolometric luminosities based on BHE and dusty QSO models differ significantly. $L_{\rm bol,~BHE}$ typically amounts to only $\sim 2\%$ of $L_{\rm bol,~QSO}$. \citet{Akins25} adopt a dusty QSO SED with $A_V \sim 3$ for LRDs, so that the strongly attenuated UV emission is reprocessed into the near-infrared, thereby elevating the inferred bolometric output. In contrast, our BHE model reproduces the optical SED with more modest dust attenuation, $A_V \sim 1$, because the blackbody continuum naturally explains the observed optical bump. The resulting bolometric correction from the rest-frame 5100~\AA\ luminosity is $\simeq 4.4$ in the BHE model, slightly smaller than the canonical value of $\sim 10$ commonly adopted for local blue quasars \citep{Richards_2006}. The combination of the reduced dust attenuation at optical wavelengths (flux suppression factor $\sim 10^{-0.4(3-1)} \simeq 0.16$) and the smaller bolometric correction for the dust-attenuation–corrected spectrum (factor $\sim [4.4 \times 10^{-0.4}]/10 \simeq 0.18$) jointly accounts for the low normalization, yielding $L_{\rm bol,~BHE} \simeq 0.02\,L_{\rm bol,~QSO}$. 
Because the BHE framework replaces the dusty-QSO interpretation in earlier work and yields substantially lower bolometric luminosities for LRDs, population-level quantities such as the luminosity function and the black hole accretion density should be revised accordingly.

The right panel of Figure \ref{comparison} shows the bolometric luminosity and photospheric temperature 
of LRDs in the Hertzsprung-Russell (HR) diagram. 
Most LRDs lie within a narrow range of $T_{\rm eff}\simeq 4500-6000~{\rm K}$, consistent with 
the stellar-atmospheric features expected in the BHE model.
This characteristic temperature range physically arises from the photospheric conditions governed by H$^-$ opacity,
as discussed for cool, inflated gaseous envelopes surrounding rapidly growing BHs \citep[e.g.,][]{Inayoshi25c,Kido25,Begelman25}.
By analogy with red giant stars on the HR diagram, we also indicate the region obtained by extrapolating the Hayashi track 
\citep{Hayashi1961} toward higher luminosities from the locus of red supergiant stars in the Milk Way.
We note that while the photospheric conditions in the BHE framework have been discussed analytically \citep{Kido25,Begelman25},
many of the detailed structural and thermal properties remain poorly constrained and will require further investigation.

The photospheric radius, i.e., the outermost layer of the envelope structure, is estimated from the luminosity and surface 
temperature through the Stefan–Boltzmann law as
\begin{equation}
    R_{\rm ph} \simeq 0.01~{\rm pc} \left(\frac{L_{\rm bol}}{10^{11}~L_\odot}\right)^{1/2}\left(\frac{T_{\rm eff}}{5000~{\rm K}}\right)^{-2}.
\end{equation}
Most LRDs have photospheric radii of $R_{\rm ph}\simeq 4\times 10^{-3}~{\rm pc}$.
A useful comparison is with the size of the broad-line region (BLR).
\cite{Bentz13} derived the empirical BLR size-luminosity relation from reverberation mapping of local AGN with 
$\lambda L_{\lambda,5100}=10^{42}–10^{46}~{\rm erg~s^{-1}}$.
Applying this relation to the typical $\lambda L_{\lambda,5100}$ value, the BLR size of LRDs is 
$R_{\rm BLR}\simeq 8\times 10^{-3}(L_{\rm bol}/10^{11}~L_\odot)^{1/2}$ pc, 
which is systematically larger than the photospheric radii of the envelope.
This size comparison suggests that the BLR would likely reside just outside the inflated envelope.
In this configuration, BLR clouds may be powered either by ionizing radiation escaping through polar funnels of 
the thick envelope structure \citep[see Figure~14 in][]{Lin25}, or by ionizing radiation produced by young stellar 
populations surrounding the BHE structure \citep[see Figure~1 in][]{Inayoshi25b}.
These possibilities can be tested through future, long-term monitoring of BLR variability.

\par
\begin{figure*}[htbp]
\centering
\begin{subfigure}{0.49\linewidth}
\includegraphics[width=\linewidth]{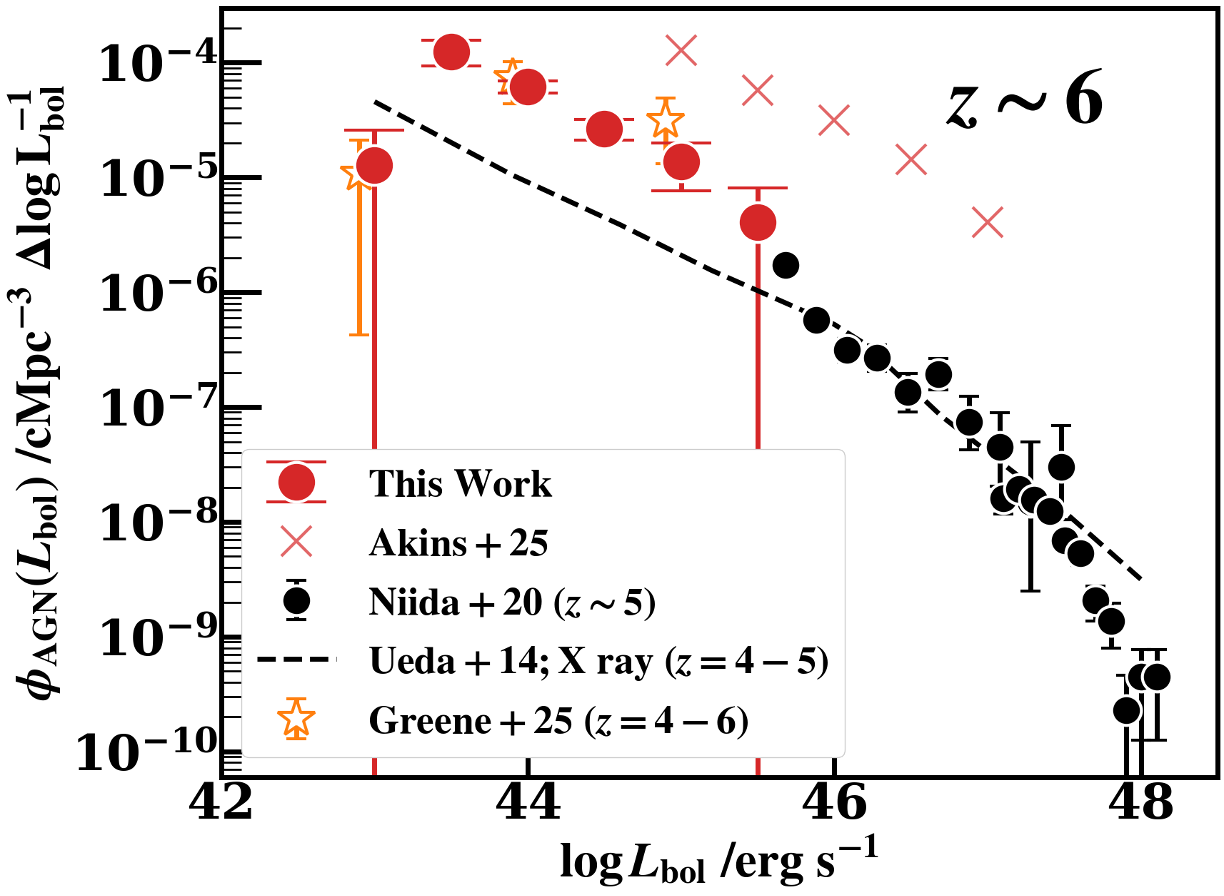}
\end{subfigure}\hfill
\begin{subfigure}{0.49\linewidth}
\includegraphics[width=\linewidth]{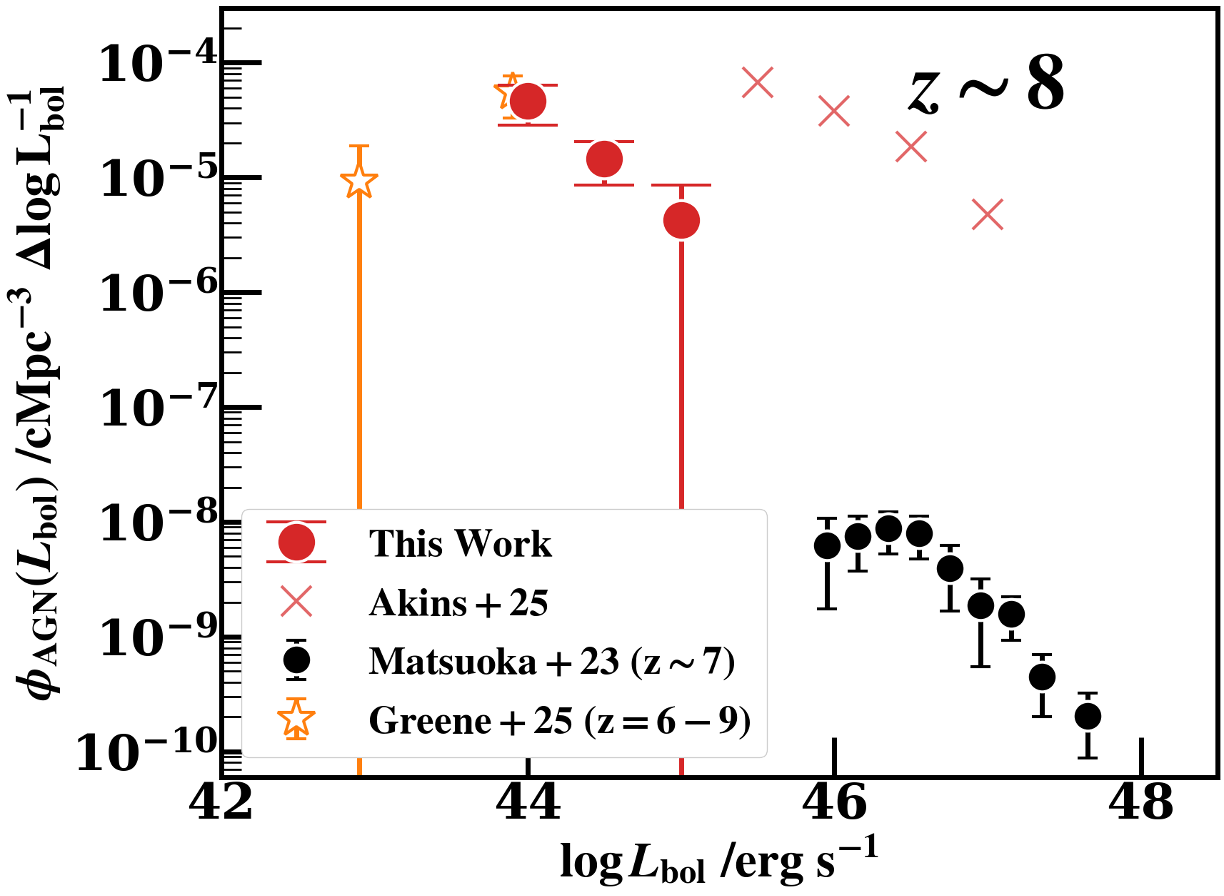}
\end{subfigure}
\caption{Left: The bolometric luminosity function of the AGNs at $z\sim6$. Number density of the LRDs are shown in the red circles. We adopt bolometric luminosity of LRD based on BHE model. We also present the bolometric luminosity function of QSO based on the Subaru surveys by \cite{Niida20}, shown in the black color. The dashed line represent the luminosity function inferred from the X-ray observations from \cite{Ueda14}. The red crosses represent the luminosity function from \cite{Akins25} derived assuming QSO model. We also show the bolometric luminosity function from \cite{Greene25} in the non-filled yellow stars. Right: The bolometric luminosity function of the AGNs at $z\sim8$. Number density of the LRDs are shown in the red color. We adopt bolometric luminosity of LRD based on BHE model. We also present the bolometric luminosity function of QSO based on the Subaru surveys by \cite{Matsuoka23}, shown in the black color. Other legends represents the same thing as they do in the left panel.}
\vspace{5mm}
\label{lf} 
\end{figure*}

\section{Revisiting Statistical Properties of LRDs}
\subsection{Luminosity Function}
Based on the inferred SED parameters, we calculate the bolometric luminosity function for LRDs. We simply integrate the blackbody part of the radiation in the equation down to the rest-frame Balmer break wavelength for the LRD bolometric luminosity. We divide the sample into redshift bins of $5<z<7$ and $7<z<9$ to calculate the AGN luminosity function at $z\sim6$ and 8, respectively. We also consider the completeness correction based on \citet{Akins25}. We calculate the comoving volume for each redshift bin based on the survey area of COSMOS-Web covered by MIRI, which is 0.19 deg$^2$. We estimate the completeness as a function of $L_{\rm bol}$ by anchoring our measurements to the luminosity function of \citet{Akins25}. First, for each $L_{\rm bol}$ bin in \citet{Akins25}, we compute the completeness as the ratio between their luminosity function and the raw number density of detected LRDs in that bin. We then map this completeness function to our $L_{\rm bol}$ bins using the nearly linear relation between the bolometric luminosities inferred by \citet{Akins25} and by our BHE modeling (Figure~\ref{comparison}). We adopt the mapped values as the completeness in our analysis.\par

We show our LRD luminosity function for $z\sim6$ and 8 in Figure \ref{lf}. For comparison, we also plot the QSO luminosity function from the literature based on the X-ray observations \citep{Ueda14} and the high-z QSO survey by the ground-based telescope \citep{Niida20,Matsuoka23}. For the rest-frame UV and X-ray selected sources, we apply calibrations from \cite{Duras20} to convert the observed UV/X-ray luminosities to the bolometric luminosities. Unlike the claims that the LRD number density is excessive, our number density estimates lie around the extrapolated faint end of the bright QSO luminosity function. The mitigation of apparent LRD overabundance mainly results from the reduction of the inferred bolometric luminosity by adopting the BHE model. 
We also show the bolometric luminosity function derived by \citet{Greene25}. They estimated a bolometric correction factor of $\simeq 5$ for the 5100~{\AA} luminosity by interpolating and extrapolating the fluxes from UV to Infrared wavelengths of high signal-to-noise LRD spectra and integrating the resulting SED. They recalculated LRD bolometric luminosity function from \cite{Greene24} based on the new correction factor. The bolometric correction factor of 4.4 and the luminosity function we derive with the BHE framework are in good agreement with the results of \citet{Greene25}.\par

\subsection{Black Hole Mass Function}
\begin{figure*}[t]
\centering
\begin{subfigure}{0.53\linewidth}
\includegraphics[height=0.75\linewidth]{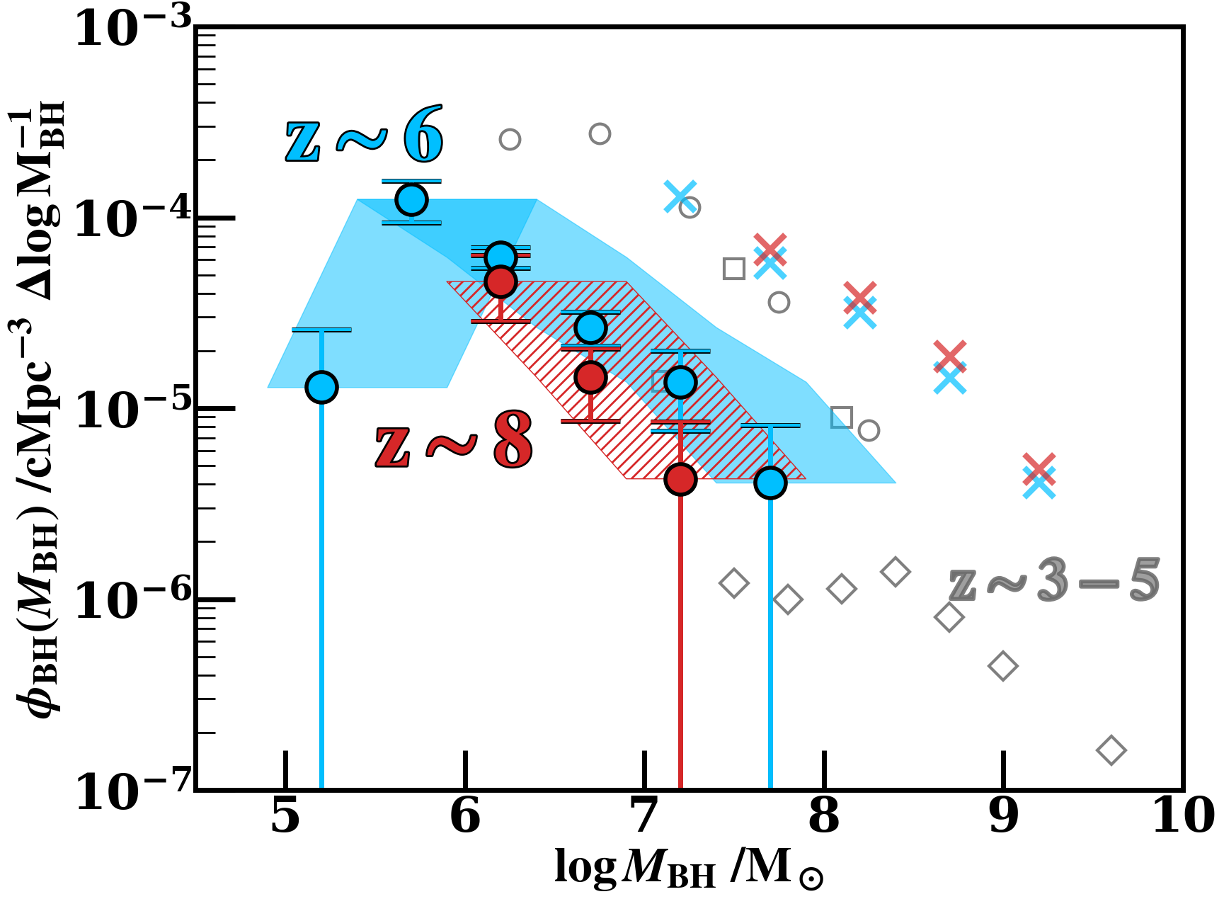}
\end{subfigure}\hfill
\begin{subfigure}{0.45\linewidth}
\includegraphics[height=0.87\linewidth]{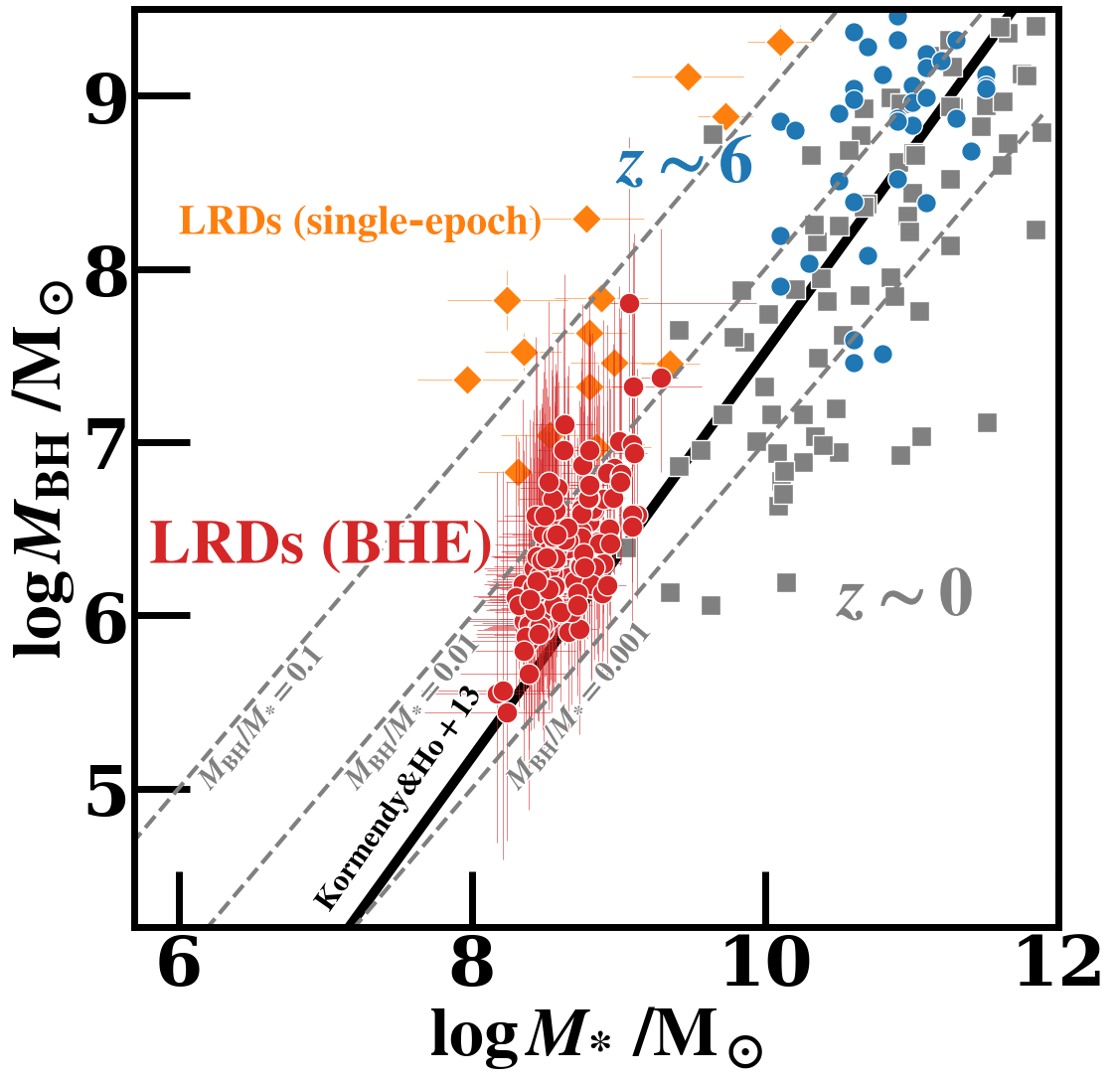}
\end{subfigure} 
\caption{Left: Black hole mass functions of LRDs at $z\sim 6$ (blue) and $z\sim 8$ (red). 
Black hole masses are estimated from the bolometric luminosity in the BHE model, assuming $\lambda_{\rm Edd}=0.5$ (filled circles). 
For each redshift, the shaded region denotes the BHMF range obtained with $\lambda_{\rm Edd}=0.1-1$.  
Cross symbols show the BHMF inferred under the dust-reddened AGN hypothesis \citep{Akins25} with $\lambda_{\rm Edd}=0.5$. 
For the comparison, BHMF constraints for AGNs at $z=3-5$ derived from single-epoch virial mass estimates are overlaid with open symbols
(circle: \citealt{2025ApJ...986..165T}, square: \citealt{Matthee24}, diamond: \citealt{2024ApJ...962..152H}).
Right: Relation between host stellar mass and black hole mass for LRDs in the BHE framework (red circles)
including uncertainties from the assumed $\lambda_{\rm Edd}$ values. 
Measurements of the $M_{\rm BH}-M_\star$ relation at $z=0$ (grey square and black diagonal line: \citealt{KormendyHo13}) and $z\sim6$ (blue circle: \citealt{Izumi21}) are shown for comparison.
The LRD population tends to be overmassive ($M_{\rm BH}/M_\star\simeq 0.005$) relative to the local relation, although the offset is moderate.
Black hole masses inferred from the H$\alpha$-based single-epoch method \citep{Kocevski25} are also plotted (orange diamonds).
Dashed lines indicate $M_{\rm BH}/M_\star = 10^{-3}$, $10^{-2}$, and $10^{-1}$.
}
\label{bhmass}
\end{figure*}

We infer the black hole mass function (BHMF) using the bolometric luminosity. The bolometric luminosity of the black hole physically relates to the mass of black hole via the following relation:
\begin{equation}
  \begin{aligned}
    M_{\rm BH}&=\frac{\sigma_T}{4\pi cG m_p}\cdot\frac{L_{\rm bol}}{\lambda_{\rm Edd}} \\
    &\quad\simeq0.81\times10^5~M_\odot \lambda_{\rm Edd}^{-1}\left(\frac{L_{\rm bol}}{10^{43}~{\rm erg~s^{-1}}}\right)
  \end{aligned}
\end{equation}
Here, $\lambda_{\rm Edd}$ denotes the Eddington ratio. 
We adopt a fiducial value of $\lambda_{\rm Edd}=0.5$ to reflect the near-Eddington accretion state expected in the BHE model. 
We also consider $\lambda_{\rm Edd} = 0.1$ and 1, and incorporate the systematic variation into the BH mass measurement uncertainties. For reference, we plot the BHMF derived from the bolometric luminosity function of \cite{Akins25} in the dust-reddened AGN hypothesis with $\lambda_{\rm Edd}=0.5$. We additionally present the black hole mass function at $z\simeq 6$ (cyan) and $z\simeq 8$ (red) in Figure \ref{bhmass}.

We further compare our results to the BHMF of AGNs at $z\sim5$ inferred from single-epoch virial mass estimates \citep[][]{2024ApJ...962..152H,Matthee24,Taylor25}. 
The BHMF derived under the BHE model lies below the $z\sim5$ virial-based BHMF. We find no strong redshift evolution between $z\sim6$ and 8. 
However, we observe an order-of-magnitude drop from $z\sim5$ to $z\simeq 6$ at fixed mass, which may partly arise from the differences in estimating the BH mass.

\subsection{Stellar Mass to Black Hole Mass Relation}
Based on our black hole mass measurements, we revisit the stellar mass and black hole mass relation based on our SED fitting results using the BHE model. According to the BHE model of \citet{Inayoshi25b}, the UV continuum radiation blueward of the Balmer break is attributed to stars. Therefore, we can derive the stellar mass from the UV luminosity assuming the scaling relation between UV luminosity and SFR and that the host galaxy lies in the SFR-$M_\star$ main sequence. For the conversion factor between UV luminosity to SFR, we assume the scaling law as follows:
\begin{equation}
  {\rm SFR}~/M_\odot~{\rm yr}^{-1}=1.15\times10^{-28}L_{\rm UV}~/{\rm erg~s^{-1}~Hz^{-1}}.
\end{equation}
\Addp{Here, we adopt a constant star formation rate and a Salpeter IMF (0.1–100~$M_\odot$), assuming stellar ages $>300$~Myr and a characteristic metallicity of $\sim0.5\,Z_\odot$, for which the results depend only weakly on metallicity \citep{Madau14}.}
For the star-forming main sequence, we adopt the empirical relation of \cite{Popesso23}. They used a sample of galaxies at $0<z<6$ spanning stellar masses
of $10^8$–$10^{12}~M_{\odot}$ to derive the following empirical calibration
between stellar mass and SFR \citep[i.e., equation 10 of ][]{Popesso23}:
\begin{equation}
  \begin{aligned}
\log {\rm SFR}/M_\odot~{\rm yr}^{-1}=&(a_1t(z)+b_1)\log M_\star/M_\odot\\
    &+b_2\log^2M_\star/M\odot\\
    &+(b_0+a_0)t(z)
  \end{aligned}
\end{equation}
Here, $t(z)$ corresponds to the age of the universe at $z$ in Gyr. We use the best-fit value for the coefficients (i.e., $a_0$, $a_1$, $b_0$, $b_1$, and $b_2$) listed in the Table 2 of \cite{Popesso23} with a correction to match to the Salpeter IMF.
In Figure \ref{bhmass}, we present our stellar and black hole mass estimate for the LRD sample. We also present the local $M_\star$-$M_{\rm BH}$ relation from \citep{KormendyHo13} and the measurements of $z\sim6$ quasars \citep{Izumi21}. We find that the majority of LRD $M_{\rm BH}/M_\star$ ratios measured with the BHE model lie on or slightly above the local relation and $z\sim6$ quasars. We compare our results with the LRD $M_{\rm BH}/M_\star$ measurements from \citet{Kocevski25}, which are derived using an empirical relation between $M_{\rm BH}$ and the broad emission lines. We see that $M_{\rm BH}/M_\star$ measurements based on BHE model are about 1–2 orders of magnitude lower than those based on the broad line relations at the fixed stellar mass. This offset aligns with the deviation of between the BHMF derived using the BHE model and that based on the empirical relation.
We note, however, that the current sample of LRDs with detected broad lines
may be biased toward overmassive black holes, because only the most luminous
broad-line systems are detectable at the survey depth. We therefore conservatively suggest that the BHE model offers a plausible way to alleviate the apparent tension between the stellar mass and black hole mass relations for LRDs. Further tests of this scenario will require larger and deeper spectroscopic samples of LRDs.

\begin{figure*}[htbp]
\centering
\begin{subfigure}{0.5\linewidth}
\includegraphics[height=0.75\linewidth]{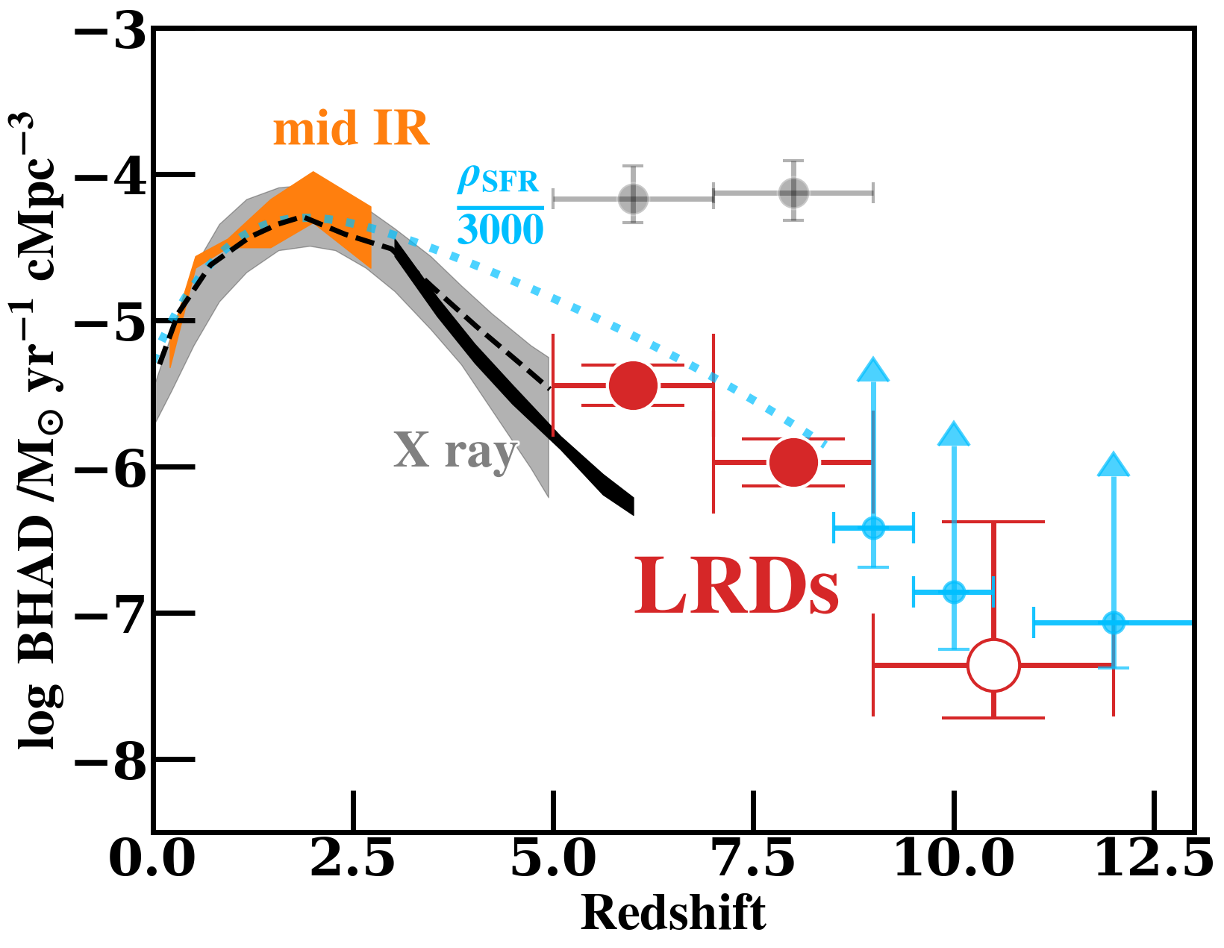}
\end{subfigure}\hfill
\begin{subfigure}{0.49\linewidth}
\includegraphics[height=0.75\linewidth]{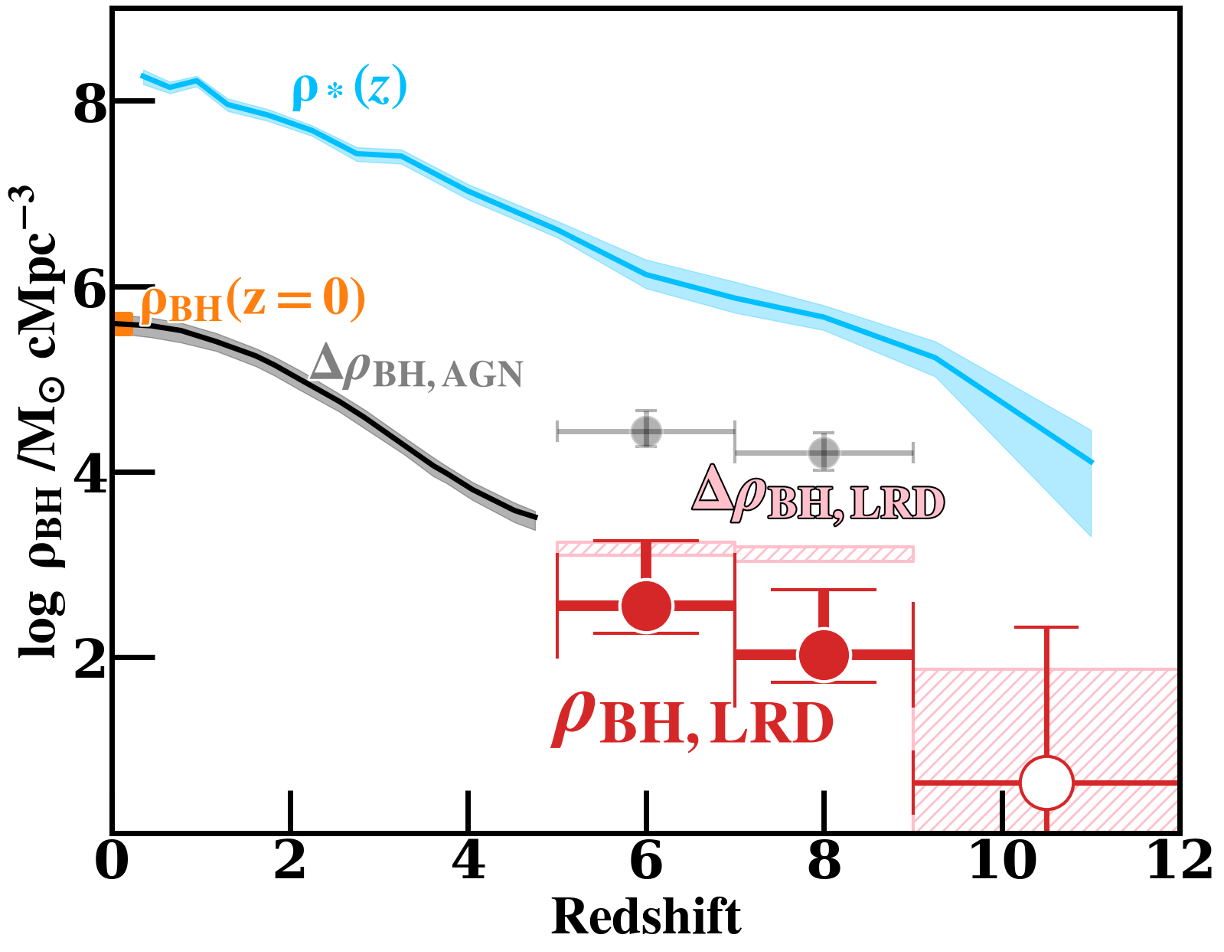}
\end{subfigure}
\caption{Left: Black hole accretion density (BHAD) of LRDs as a function of by redshift (red circles). 
BHAD are estimated from the bolometric luminosity function based on BHE model, assuming $\epsilon_{\rm rad}=0.1$. Measurements based on \cite{Akins25} $z\sim6-8$ LRD sample (filled) and $z\sim10$ LRD (open) from \cite{Tanaka25} are shown together with original BHAD measurements from \cite{Akins25} with QSO model (grey circles)
For the comparison we present BHAD constraints from observations in mid-IR (orange: \citealt{Delvecchio14}) and X-ray (grey region: \citealt{Aird15}, black region: \citealt{Pouliasis24}, black dashed line: \citealt{Ananna19}). We also plot the redshift evolution of star formation rate density $\rho_{\rm SFR}$ in blue symbols (dotted line: \citealt{Harikane22},  circles:\citealt{Harikane25a}). We scale $\rho_{\rm SFR}$ by a factor of 1/3000 to match the BHAD at $z<3$. 
Right: The black hole mass density (BHD) by redshifts. The red circles represent the $\rho_{\rm BH}$ derived from integrating BHMF of LRDs at $z=6$ and 8. The open circle represent the $\rho_{\rm BH}$ estimated using a $z>10$ LRD candidate reported by \cite{Tanaka25}. The error bars include the systematic offset arising from adopting $\lambda_{\rm Edd}=0.1$ versus $\lambda_{\rm Edd}=1$. The pink shades represents the $1\sigma$ upper and lower limit ranges $\Delta\rho_{\rm BH}$ calculated from integrating BHAD of LRD by time ($\Delta t$) corresponding to the redshift binning width. The grey circles represent the $\Delta\rho_{\rm BH}$ estimated based on BHAD assuming QSO model \citep{Akins25}.
The orange region represent the $\rho_{\rm BH}$ at $z=0$ measured by \cite{Shankar09}. The black solid line represent the $\Delta\rho_{\rm BH}$ estimated from the BHAD of $0<z<5$ ordinary AGNs with $\epsilon_{\rm rad}=0.1$ \citep{Ueda14}. 
For the comparison, we plot recent stellar mass density measurements from \cite{Shuntov25} in blue line.}
\vspace{5mm}
\label{bhad_bhd} 
\end{figure*}

\subsection{Black Hole Accretion Density}
Next, we calculate the black hole accretion density for the LRDs using the inferred bolometric luminosity from the BHE model. The mass accretion rate of the black hole $\dot{M}_{\rm BH}$ can be expressed in terms of bolometric luminosity in the following manner:
\begin{equation}
\dot{M}_{\rm BH} = \frac{1-\epsilon_{\rm rad}}{\epsilon_{\rm rad}} \cdot \frac{L_{\rm bol}}{c^2},
  \label{bhad_eq}
\end{equation}
where $\epsilon_{\rm rad}$ represents the radiative efficiency of the accreting black hole. We adopt a fiducial value of $\epsilon_{\rm rad}=0.1$ as commonly assumed in the literature \citep[e.g.,][]{Soltan82,Yu02,Ueda14}. 
The radiative efficiency may be reduced when the BH accretes at rates exceeding the Eddington limit as expected in the BHE model.
Radiation-hydrodynamic simulations show that even when gas is supplied at highly super-Eddington rates on large scales, convection within the envelope suppresses the net inflow toward the BH, so that the BH is actually fed at only mildly super-Eddington rates of $\dot{M}_{\rm BH}=\mathcal{O}(1-10)\times \dot{M}_{\rm Edd}$.
Under these conditions, the radiative efficiency decreases by a factor of $\simeq 2$ relative to the canonical value \citep[e.g.,][]{Jiang19,Inayoshi20}.
Despite the uncertainty in $\epsilon_{\rm rad}$, we adopt $\epsilon_{\rm rad}=0.1$ as our baseline value to maintain consistency with the literature and to allow direct comparison, while noting that moderate deviations fro this choice would not alter our main conclusion.

We then estimate the black hole accretion rate density (BHAD) traced by the observed LRDs by integrating their luminosity function down to the COSMOS-Web detection limit and converting the bolometric luminosity density into mass accretion density using equation \ref{bhad_eq}. 
We present our BHAD measurements in Figure \ref{bhad_bhd} with the BHAD constrained from literature using mid-IR \citep{Delvecchio14} and X-ray observations \citep{Aird15,Pouliasis24,Ananna19}. We find that the BHAD of LRDs at $z \sim 6$–8 lies on the extrapolated high-redshift tail of the BHAD constraints from the lower redshift observations. For the reference, we also calculate the BHAD contributed by an LRD-phase black hole at $z > 10$ LRD candidate \citep{Tanaka25}. We adopt the estimate number density and the bolometric luminosity (i.e., $L_{\rm bol}\sim3\times10^{44}~{\rm erg~s^{-1}}$) inferred by \cite{Tanaka25} with BHE model. Reduction of the inferred luminosity using the BHE model naturally reconciles the overshooting BHAD values inferred from the ordinary QSO template models. Moreover, the smooth connection between the BHAD inferred at $z<5$ from X-rays and at $z>5$ from LRDs suggests that the LRD-phase may have played a dominant role in the black hole mass accumulation at the epoch of cosmic reionization. The gradual transition from the LRD-dominated to the X-ray–emitting, normal AGN population through super-Eddington accretion around $z \sim 5$ may possibly mark the emergence of ordinary quasars after LRD-phase.

\subsection{Black Hole Mass Density Evolution}
According to Soltan's argument, the cumulative mass accreted onto black holes should match the total black hole mass density in the local universe. 
In Figure \ref{bhad_bhd}, we present present the black hole mass density of LRDs ($\rho_{\rm BH}$) inferred from their observed abundance, along with the increase in their mass density ($\Delta\rho_{\rm BH}$) obtained by integrating the BHAD over the time interval of each redshift bin (i.e., $\Delta\rho_{\rm BH}\simeq{\rm BHAD}\times\Delta t$).
We find that the $\Delta\rho_{\rm BH}$ contributed by LRDs is smaller by several 
factors than the total BH mass growth at the corresponding redshifts.
The comparison between the observed $\rho_{\rm BH}$ of LRDs at $z\sim6$ and 
the cumulative $\Delta\rho_{\rm BH}$ produced during the LRD phase from $z=10$ to $z=6$
suggests that the LRD phase has a duty cycle of about 20\%. 
This inferred value is consistent with the duty cycle estimated from clustering analyses
\citep[i.e., $\sim40\%$;][]{Arita25}, although other clustering-based studies indicate 
even lower values \citep[e.g., $\sim1\%$;][]{2025arXiv250502896L}.

To check the consistency between the high and low redshift BH populations, we present the observationally constrained $\rho_{\rm BH}$ at the local universe from \cite{Shankar09} together with $\Delta\rho_{\rm BH}$ calculated from the AGN BHAD at $0<z<5$ assuming $\epsilon_{\rm rad}=0.1$ \citep{Ueda14}. The cumulative mass accretion from the ordinary AGN population at $0<z<5$ smoothly connect the $z=0$ $\rho_{\rm BH}$ and $z>6$ $\Delta\rho_{\rm BH}$ from LRDs, implying that mass accretion with a radiative efficiency of $\epsilon_{\rm rad} \simeq 0.1$ is consistent with Soltan's argument over $0 \lesssim z \lesssim 10$ \citep[see also][]{Zou2024}.\par 
In Figure \ref{bhad_bhd}, we also plot redshift evolutions of star formation rate density measurements from \cite{Harikane22} and \cite{Harikane25a} along with BHAD evolution, and stellar mass density measurements from \cite{Shuntov25} along with black hole mass density evolution. With a normalization of SFR density by a factor of 3000, the redshift evolution of BHAD and SFR density coincide with one another. Moreover, we find that the stellar and black hole mass densities evolve in parallel with redshift up to $z\sim10$. 
Moreover, the parallel evolution between stellar and black hole mass densities at $0<z<10$ could possibly indicate the coevolution of black hole and host galaxies from the early stage of black hole formation. Such coevolution may be driven by linked mass assembly, in which efficient cold-stream gas inflows feed the central regions of dark matter halos \citep[e.g.][]{2006MNRAS.368....2D}, while subsequent supernova feedback from the young nuclear stellar cluster disrupts the LRD phase \citep{Inayoshi25b}. Together, these processes could possibly provide a natural pathway for the concurrent build-up of black hole and stellar mass, ultimately giving rise to the tight black hole–stellar mass correlations observed in the local Universe.
\section{Conclusion}\label{sec:conc}
In this Letter, we reanalyzed the COSMOS-Web LRD sample using the black-hole-envelope (BHE) model, which describes super-Eddington accretion surrounded by an optically thick envelope. The BHE model reproduces the characteristic optical V-shaped continuum through photospheric emission at $T_{\rm ph
}\sim 4000-6000$ K and naturally explains the lack of far-infrared detections without invoking heavy dust obscuration. The inferred bolometric luminosities are typically 1-2 dex lower than those derived from quasar templates, bringing the number densities of LRDs into agreement with the faint-end extrapolation of known AGN luminosity functions. The resulting black hole mass function and accretion-rate density smoothly connect to those of ordinary AGNs at $z\lesssim5$, suggesting that LRDs represent a short-lived, efficient growth phase in the early build-up of supermassive black holes.
Expanded MIRI coverage from programs such as COSMOS-3D \citep[GO-5893; PI. Koki Kakiichi; ][]{2024jwst.prop.5893K} will improve constraints on the rest-frame optical–to–NIR spectral properties of LRDs.
Future spectroscopy (e.g. EMBER program; GO-7076; PI: Hollis Akins) and variability monitoring will be crucial to confirm the physical nature and duty cycle of this LRD-phase.
\section*{Acknowledgments}\label{ack}
We thank Hollis Akins and COSMOS-Web teams led by Jeyhan Kartaltepe for public data release of COSMOS-Web and LRD catalog. We thank Masami Ouchi, Kunihito Ioka, Tatsuya Matsumoto, and Hiroto Yanagisawa for the useful comments and discussions. This work is based on observations with the NASA/ESA/CSA James Webb Space Telescopes at the Space Telescope Science Institute, which is operated by the Association of Universities for Research in Astronomy, Inc., under NASA contract NAS 5-03127 for JWST. 
H.U. acknowledges support from the World Premier International Research Center Initiative (WPI Initiative), MEXT, Japan, and KAKENHI (20H00180) through Japan Society for the Promotion of Science, the joint research program of the Institute for Cosmic Ray Research (ICRR), University of Tokyo, and FoPM WINGS Program in the University of Tokyo.
K.I. acknowledges support from the National Natural Science Foundation of China (12573015, 1251101148, 12233001), 
the Beijing Natural Science Foundation (IS25003), and the China Manned Space Program (CMS-CSST-2025-A09).
Y.H. acknowledges support from the JSPS Grant-in-Aid for Scientific Research (24H00245) and the JSPS International Leading Research (22K21349).
The work of K.M. was supported by KAKENHI No.~20H05852 and the Institute for Gravitation and the Cosmos of the Pennsylvania State University.
We thank the Yukawa Institute for Theoretical Physics at Kyoto University. Discussions during the YITP workshop YITP-W-25-08 on Exploring Extreme Transients were useful to complete this work.


\bibliography{main}{}
\bibliographystyle{aasjournalv7}



\end{document}